\documentclass[12pt,a4paper]{article}
\usepackage{epsfig}
\title{Shell Structure of Cesium Layer Covering the C$_{60}$ Fullerene Core
\thanks{This work has been partly supported by the Polish Committee for 
        Scientific Research under Contract No. 2P03B 115 19 and by the Program 
        of Scientific Exchange between the IN2P3-France and the Polish 
        Research Institution No. 99-95}}

\author{          P. Mierzy\'nski$^1$, K. Pomorski$^{1,2}$\\
{\it $^1$Katedra Fizyki Teoretycznej, Uniwersytet M. C. S.,
                               Lublin, Poland}
\\
      {\it $^2$ Institute de Recherches Subatomiques (IN22P3 -- CNRS)}\\ 
      {\it and Universit\'e Louis Pasteur, Strasbourg, France}} 

\begin{document}

\maketitle

\begin{abstract}
\noindent
Strutinsky shell corrections for the Cesium-coated fullerenes were
investigated. The single particle levels of electrons are obtained using the
spherical mean-field potential of a shifted Wood-Saxon type. The parameters of
the potential are adjusted to reproduce the experimental ionization energies 
of the Cs($N$) clusters and the magic numbers observed in their 
photo-ionization spectra of the C$_{60}$Cs($N$) aggregates. \\

Electronic structure of clusters and nanoparticles

\end{abstract}

\section {Introduction} 

The world wide interest on metallic clusters started in the early 80's when a
group of physicists from Berkeley performed the first experiments with tiny 
alkali clusters \cite{Kni84,Kni85}. Since that time an enormous amount of work
both on the experimental and theoretical side has been accomplished. Metallic
clusters are the quantum many body systems built from several up to a few
thousands atoms and their radii are of the order of 10 {\AA}. It is well
known that the shell structure of electronic orbitals in the metallic cluster
can be well reproduced not only by selfconsistent calculations but also within
some average potential approach based on the jellium model (see e.g.
\cite{Bra93}). In the jellium model one assumes, that electrons are moving in
the mean field potential which is produced by the background of positive ions
and by other free electrons.

In 1996 an experimental group from Stuttgart \cite{Fra97} succeeded in
producing and the measuring optical response of Cesium coated fullerenes
C$_{60}$ covered by a layer of $N$ Cs-atoms with $N\leq 500$. The electronic
shell structure of these C$_{60}$Cs($N$) agglomerates was studied in
Ref.~\cite{Spr96} using the local approximation to the density-functional
theory. The measured abundances of metal-coated fullerenes C$_{60}$Cs($N$) were
interpreted there in terms of the magic numbers corresponding to the electronic
shell closures. Recently, in Ref.~\cite{Pom98} the shell structure of thin
spherical metallic layers was discussed using the Strutinsky shell correction
method \cite{Str66} and an infinite square well approximation to the mean field
potential for electrons. In spite of this rather rough approximation to the
average mean field potential it was shown there that for C$_{60}$Cs($N$) the
minima of the shell correction as function of the atom number $N$ in the
cluster were found not far from the experimental magic numbers given in
Ref.~\cite{Spr96}. The agreement was better at higher mass numbers, where the
jellium model is more reliable. 

The aim of the present investigation is to use a similar approach as in 
Ref.~\cite{Pom98} but with a more realistic approximation to the mean-field
potential. We are going to keep the spherical symmetry of the potential
because the fullerene C$_{60}$ core is a spherical ball, what additionally
stabilizes the system against deformation. We show in the following, that a
shifted Saxon-Woods potential reproduces reasonably well the experimental magic
numbers and the ionization energies. It will also be shown how the magic
numbers of the Cesium layer change with the core radius. This is not
only an {\it academic} study since one can easily imagine an experiment in
which one covers a spherical insulator with a thin metal layer.


\section{Theoretical model}

The valence electrons have a decisive influence on the abundance of metallic
clusters. The electrons are moving in a mean field potential which is produced
by the jellium, i.e. the positively charged ions and the rest of the electrons.

It was shown (see discussion in Ref.~\cite{Bra93}) that the Saxon-Woods type 
potential
\begin{equation}
 V_{phen}(r)= {-V_0\over1+e^{r-R_0\over a}}~,
\label{wsax}
\end{equation} 
approximates quite well the mean field which is felt by free (valence) 
electrons. The radius $R_0$ in (\ref{wsax}) has to be chosen as
\begin{equation}
  R_0= r_0N^{1/3} \,\,, 
\label{crad}
\end{equation}
with
\begin{equation}
  r_0= r_{00}(1+{\alpha \over N^{1/3}}) \,\,, 
\label{crad1}
\end{equation}
where $ r_{00}$ is the Wigner-Seitz cell radius, $N$ the number of atoms in 
the cluster and $\alpha$ a constant originating from the so-called 
"spilled-out" effect \cite{Kni84,Bra93}. The potential depth $V_0$ and the
surface thickness $a$ are usually adjusted to the experimental values of
ionization energies and to some spectroscopic data (e.g. magic numbers) of
clusters in a broad range of $N$. Solving the Schr\"odinger equation
\begin{equation} 
 \left(-{\hbar^2\over2m_e}\nabla^2 + V_{phen}\right)\Psi_\nu=
    e_\nu\Psi_\nu \,\,,
\label{ham}
\end{equation} 
one can obtain the ionization energy by looking for the energy of the last
occupied level (Fermi energy). One has to keep in mind that due to the spin of
the electron one has to put two electrons on each single-particle level $e_\nu$.
In the spherical case the levels are strongly degenerated
$$\Psi_\nu = \Psi_{n\ell m m_s},~~~~ {\rm while} ~~~~ e_\nu = e_{n\ell}\,\,.$$ 
The magic numbers can be roughly identified with the numbers of particles 
below energy gaps in the single-particle spectrum $e_\nu$ or more accurately 
as minima in the considered as function of the particle number $N$ Strutinsky 
shell correction $\delta E_{shell}(N)$ \cite{Str66,Nil69}
\begin{equation}
\delta E_{shell}(N) = 2\sum_{occ}e_{\nu}-\widetilde {E}(N) \,\,,
\label{str}
\end{equation} 
where $\widetilde {E}$ is energy of electrons in the cluster with smeared-out 
shell structure. We have used here, as usually done in the Strutinsky procedure,
a Gauss function with the six order correctional polynomial to washed out 
the shell effects. The width ($\gamma$) of the Gaussian was fixed from the 
plateau condition of
\begin{equation}
\widetilde {E}(N)= 2\sum_{\nu=1}^\infty e_\nu \tilde n_\nu(\gamma) \,\,,
\end{equation} 
where $\tilde n_\nu$ is the average occupation number of the level $e_\nu$.

We are going to discuss the shell structure of the electron orbitals in thin 
Cesium layers built on a fullerene core or another spherical ball made from 
insulator. A shifted Saxon-Woods potential seems to be most appropriate in this 
case. Such a potential, shown in Fig.~1, is described by the following equation 
\begin{equation}
V(r) = V_0 \left({1\over 1+e^{r-R_2\over a_2}}
     -{1\over 1+e^{r-R_1\over a_1 }}\right) \,\,,
\label{swsax}  
\end{equation}
where $ R_2$ is the radius of the inner core and $R_1$ depends on the number of 
atoms in the cluster
\begin{equation}
  R_1= (r_0^3N+R_2^3)^{1/3}.
\label{agrad}
\end{equation}
The parameters $a_1$ and $a_2$ describe the diffuseness of the outer and inner
potential wall at $R_1$ and $R_2$ respectively. The parameter $a_1$ should be
essentially equal to the parameter $a$ from Eq.~(\ref{wsax}) for the compacted
cluster. The potential which feels the electron in the region between the core 
and the metallic layer of the Cesium atoms is described by the parameter
$a_2$ which will be adjusted to reproduce the observed magic numbers in the 
C$_{60}$Cs($N$) agglomerates as explained in the next section.


\section{Results}

The single-particle levels of electrons were evaluated using the Woods-Saxon
potential (\ref{wsax}) for the compact spherical clusters build from a few to 
1000 atoms. The magnitude of the ionization energy dependens on the depth 
$V_0$ of the potential, its radius $R_0$, and surface thickness $a$:  
\begin{equation}
  E_{io} = e_{N/2}(V_0,R_0,a) \,\,.
\label{eio}
\end{equation}

In order to fix the parameter set of the potential (\ref{wsax}) we have
performed a preliminary computation for Sodium Na($N$) and Potassium K($N$)
clusters which are the best known clusters of metals with one valence electron
and for which the ionization energies were measured in Refs.~\cite{Hee93},
in a relatively broad range of $N$. 

We have found that using the radius shift constant $\alpha=0.8$ and the depth
of potential $V_0=6.0$ eV for Sodium and 4.6 eV for Potassium clusters, and the
surface thickness $a$=0.3~{\AA} for both type of clusters one can reproduce
quite accurately the experimental systematics of ionization energies as well as
their average phenomenological dependence on $N$ \cite{Tam99}: 
\begin{equation}
  E^{phen}_{io}(N) = {3\over 8} {e^2\over r_0 N^{1/3}} - E_{io}(\infty) \,\,,
\label{eiop}
\end{equation}
where $E_{io}(\infty)$ is the ionization energy equal to 2.7 eV for Sodium
and 2.4 eV for Potassium metals. 

The comparison of the theoretical and the experimental ionization energies and
their average dependences on $N$ is presented in Fig.~2 for the Sodium clusters
and in Fig.3 for the Potassium clusters. The experimental magic numbers for 
Sodium and Potassium as well as the other alkaline clusters are the same and 
equal
\begin{equation}
 N_{exp} =2,8,18,20,34,40,58,92,138,198,264,344,442\,\,.
\label{expmn}
\end{equation}
The theoretical magic numbers closest to the experimental ones 
\begin{equation}
 N_{th} =2,8,18,20,34,40,58,92,138,186,198,254,338,440\,\,,
\label{thmn}
\end{equation}
which are obtained with the diffuseness parameter of the Saxon-Woods potential 
equal to $a$=0.3 {\AA}.

The Cesium with the ionization energy $E_{io}(\infty)$=3.89 eV belongs also to
the group of the alkaline metals, so the Cs clusters should have similar
properties as the Na and K clusters. It is the main reason why we have adopted 
the same value of the radius shift parameter (\ref{crad1}) $\alpha=0.8$ 
performing calculations for the Cs aggregates. The theoretical systematics of 
the ionization energies for the Cs($N$) clusters obtained with the depth of the
Saxon-Woods potential equal to $V_0=5.6$ eV and with the surface thickness
$a$=0.3 {\AA} is very close to its phenomenologic estimate~\cite{Tam99} as it
can be seen in Fig.~4. If the surface thickness $a$ is larger than 0.5 {\AA}
the asymptotic behavior of the ionization energy differs significantly from the
estimate made in Ref.~\cite{Tam99}. In Fig~5 the Strutinsky shell correction
for the compact spherical Cesium clusters is plotted. The minima of the shell
correction (\ref{str}) correspond to the predicted magic numbers. It is seen
that the theoretical magic numbers are exactly the same as the experimental
ones for $N\leq 138$ and very close to the measured ones for heavier
clusters. 

Varying the inner and outer surface widths of the shifted Saxon-Woods
potential (\ref{wsax}) and keeping the rest of the parameters constant we have
tried to find a combination of the $a_1$ and $a_2$ widths leading to the magic
numbers experimentally found in Ref.~\cite{Fra97} for the C$_{60}$Cs($N$)
aggregates. The experimental and theoretical magic numbers from
Ref.~\cite{Spr96} obtained in the local density-functional theory are shown in
the first two columns of Table~1. The magic numbers obtained in
Ref.~\cite{Pom98} for the square well potential are printed in the third
column. In Columns 4-9 we have displayed the magic numbers obtained with the
potential (\ref{swsax}) for different sets of the surface thicknesses. We begin
with the outer surface thickness $a_1$ significantly larger than the inner one
($a_2$) in Column 4 and we end with an opposite choice of the thicknesses in
Column 9. It is seen that, in spite of the dramatic change of the parameters,
the predicted magic numbers for the lighter clusters do not differ
significantly. More pronounced differences are found for the middle mass and
heavy Cs clusters. We follow the suggestion made in Ref. 5 when evaluating the
positions of the magic numbers in C$_{60}$Cs($N$) and assume that 6 electrons
are accumulated in the fullerene core. 

There is no reason why the parameter ${a_1}$ in Eq.~(\ref{swsax}) for
C$_{60}$Cs($N$) should not be equal to the parameter $a$ in Eq.~(\ref{wsax})
for the compact cluster. So, in the following, we have restricted our model
and we keep $a_1=0.3$ {\AA} and we change only the inner surface thickness. The
magic numbers closest to the experimental data obtained with the above
restriction are shown in the last column of Table 1. An example of evaluated
shell corrections for C$_{60}$Cs($N$) clusters is plotted in Fig.~6. The minima
of $\delta E_{shell}$ correspond to the magic numbers predicted by our model
when one take into account the electrons being inside the C$_{60}$ core. One
can observe some differences in the positions of the minima in comparison with
Fig.~5. This is due to the effect of the inner potential wall originating from the
fullerene core. The amplitude of the shell correction reaches 1 eV for the
heavier C$_{60}$Cs($N$) clusters.

The electron single particle levels in the C$_{60}$Cs($N$) aggregates are
plotted in Fig.~7 as function of the thickness of the Cs layer which is equal
to the difference between outer radius $R_1$ and the radius of the fullerene
core $R_2$. The positions of the Fermi level for different cluster sizes are
marked by the black points. It is seen in Fig.~7 that for the large thickness
of the Cesium layer the effect of the repulsive core in the single particle
potential (\ref{swsax}) becomes negligible while for the smaller thicknesses
the level corresponding to the low angular momentum are pushed up. This figure
is very similar to the single particle plots presented in Ref.~\cite{Pom98}
for the metallic bubbles.

\section{Summary}

The phenomenological Woods-Saxon potential with an repulsive inner core
describes relatively well the values of the magic numbers observed in the
abundance of the C$_{60}$Cs(N) clusters. We have shown how strongly the
structure of the single-particle electron levels changes with the thickness of
the metal layer covering the spherical core.

The results of our investigation could be also important for practical or even
technological applications, because we indicate how one can tune in an
almost continuous way the eigenenergies of electrons. In other words we give a
prescription how to change the optical response of thin metallic layers 
covering arbitrary spherical cores made e.g. from insulator or other material
with significantly different properties, by changing the thickness of the
layer or the core size.

\newpage
~~~~~~~~~~~~~~~~~~
\vskip 40 pt

{\bf Table 1.} Comparison between the magic numbers for several theoretical and 
expe\-ri\-men\-tal results. The single particle potential thickness parameters 
$a_1$ and $a_2$ are in  {\AA} units.\\

\begin{center}
\begin{tabular}{|l|p{0.7 cm}|c|c|c|c|c|c|c|c|c|}
\hline
 &  &  &$a_1,a_2$&$a_1,a_2$&$a_1,a_2$&$a_1,a_2$&$a_1,a_2$&$a_1,a_2$&$a_1,a_2$\\
\hline
Exp.\cite{Fra97}& \cite{Spr96}&\cite{Pom98}&0,1,1.6&0.4,1.2&0.8,0.8&1.2,0.4
                                                   &1,60.1&2.0,0.1&0.3,0.1 \\
\hline
   ~~~1   &~~2  &    3  &    4  &   5 &   6 &   7  &     8 &    9  &     10 \\
\hline\hline
 12$\pm$0 &  14 &   14  &   14  &  14 &  14 &  14  &    14 &   14  &     14 \\
\hline
 27$\pm$1 &  24 &   24  &   24  &  24 &  24 &  24  &    24 &    24 &     24 \\
\hline
 33$\pm$1 &  38 &   38  &   38  &  38 &  38 &  38  &    38 &    38 &     38 \\
\hline
 44$\pm$1 &     &       &       &     &     &      &       &       &        \\
\hline
 61$\pm$1 &58,64&   58  &   56  &  56 &  64 &  64  &    64 &    64 &     64 \\
\hline
 98$\pm$1 &  96 & 86,96 &   86  &86,92&  96 &  96  &    96 & 96,110&     96 \\
\hline
146$\pm$2 & 136 &  136  &122,136& 136 & 136 & 136  &   138 &138,162&    136 \\
\hline
198$\pm$0 & 192 &186,192&  184  & 186 & 192 & 192  &   202 &   202 &    192 \\
\hline
255$\pm$5 & 258 &  ~258 &248,258& 258 & 258 & 258  &258,272&   272 &    258 \\
\hline
352$\pm$10& 338 &  ~336 &  336  & 336 & 336 & 338  &338,362&   362 &338,344 \\
\hline
445$\pm$10& 434 &  ~434 &  428  & 434 & 434 & 444  &444,466&444,466&    444 \\
\hline
\end{tabular}
\end{center}

\vspace{2cm}
\begin{center}
{\bf Figure captions:}
\end{center}

\begin{enumerate} 
\item The shape of the spherical average potential for electrons in 
      the C$_{60}$Cs($N$) cluster.
\item  The ionization energy of Sodium clusters Na($N$).  
\item  The ionization energy of Potassium clusters K($N$).  
\item  The ionization energy of Cesium clusters Cs($N$).  
\item  The shell-correction energy and the magic numbers for the 
       Cs($N$) cluster evaluated with the surface thickness 
       $a$=0.3 {\AA}. 
\item   Shell-correction energy and the magic numbers for the 
        C$_{60}$Cs($N$) cluster evaluated with the surface thickness 
        $a_1$=0.3 {\AA} and $a_2$=0.1 {\AA}.
\item The single-particle levels in the Cesium layer covering the spherical
      core of radius $R_2$. The depth of the single particle potential
      is equal to 5.6 eV, and the surface thicknesses $a_1$=0.3 {\AA} and 
      $a_2$=0.1 {\AA}.  The black points show the position of the Fermi level. 
\end{enumerate}

\pagebreak[5]
\begin{figure}
\epsfxsize=160mm \epsfbox{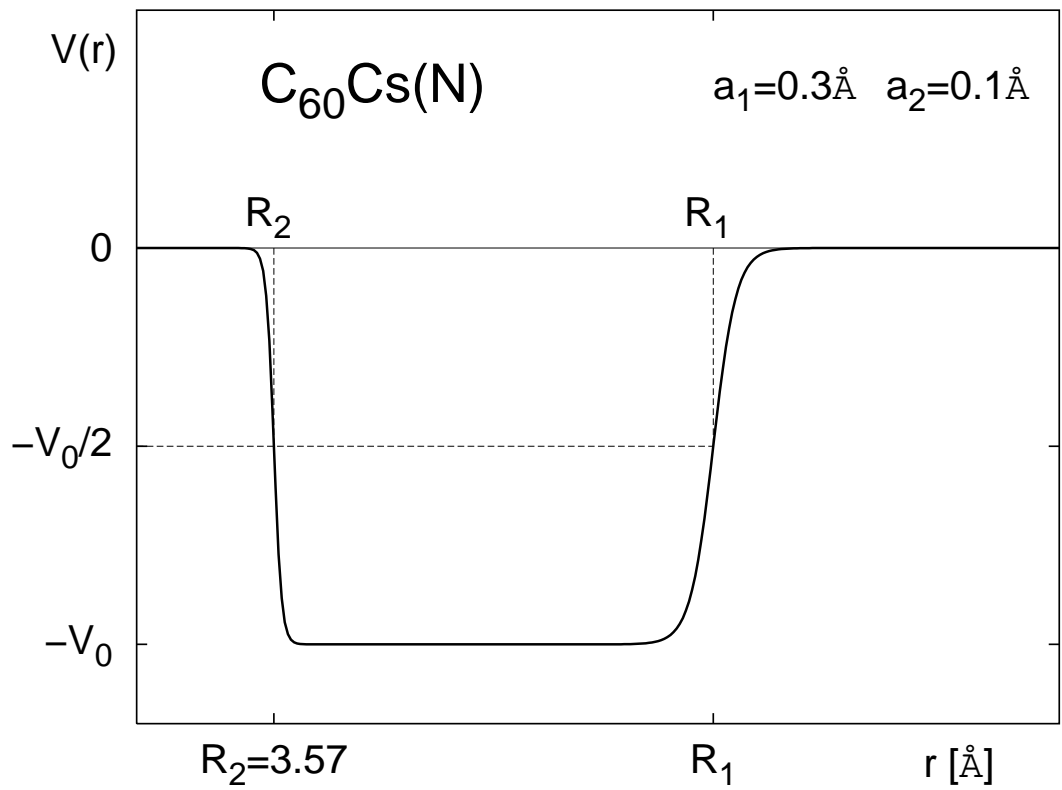}
\caption{The shape of the spherical average potential for electrons in 
      the C$_{60}$Cs($N$) cluster.}
\end {figure}

\pagebreak[5]
\begin{figure}
\epsfxsize=160mm \epsfbox{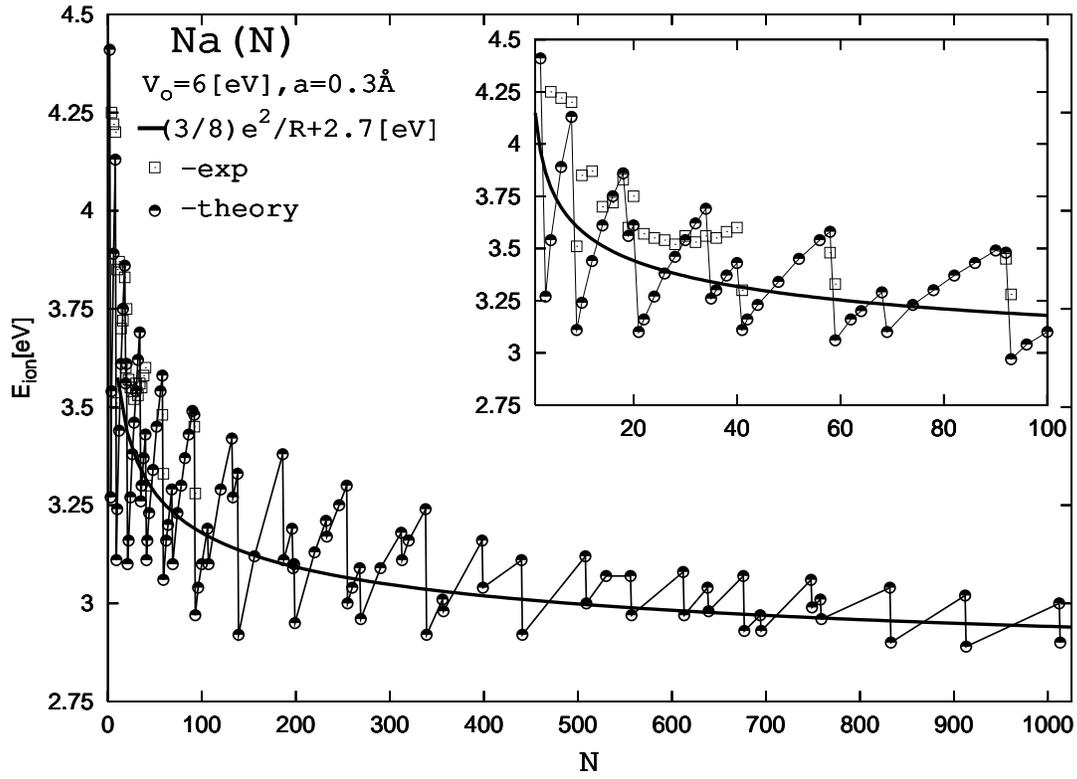}
\caption{The ionization energy of Sodium clusters Na($N$).}
\end{figure}

\pagebreak[5]
\begin{figure}
\epsfxsize=160mm \epsfbox{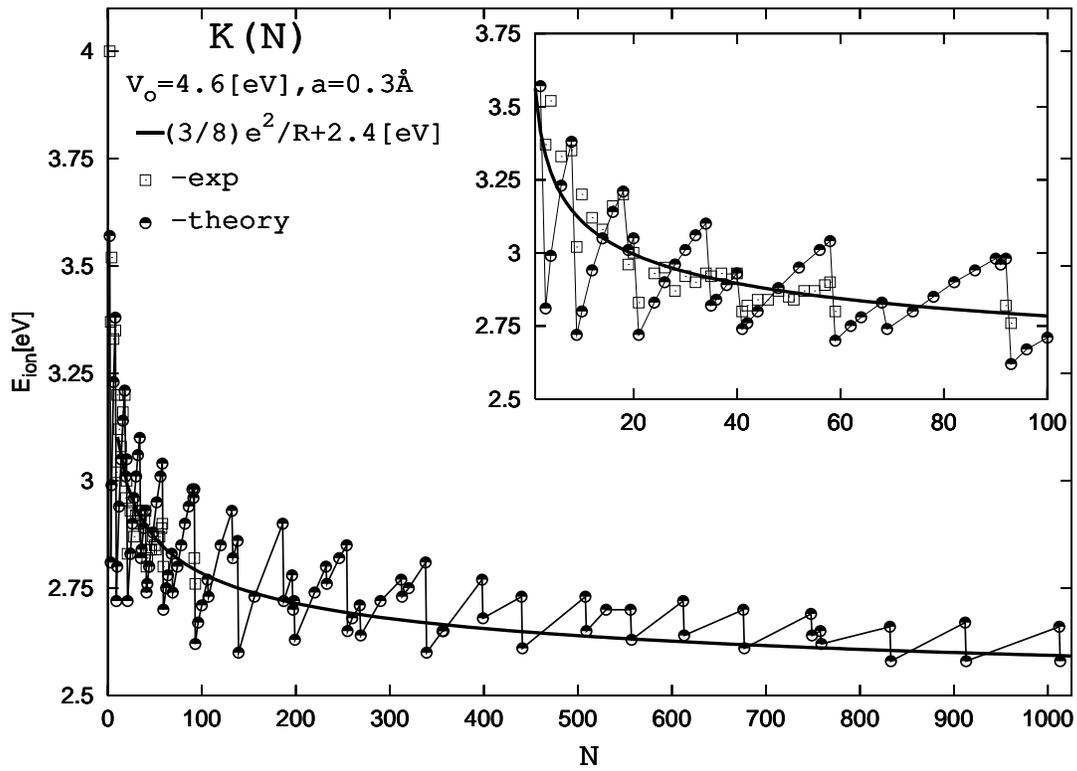}
\caption{The ionization energy of Potassium clusters K($N$).}
\end{figure}

\pagebreak[5]
\begin{figure}
\epsfxsize=160mm \epsfbox{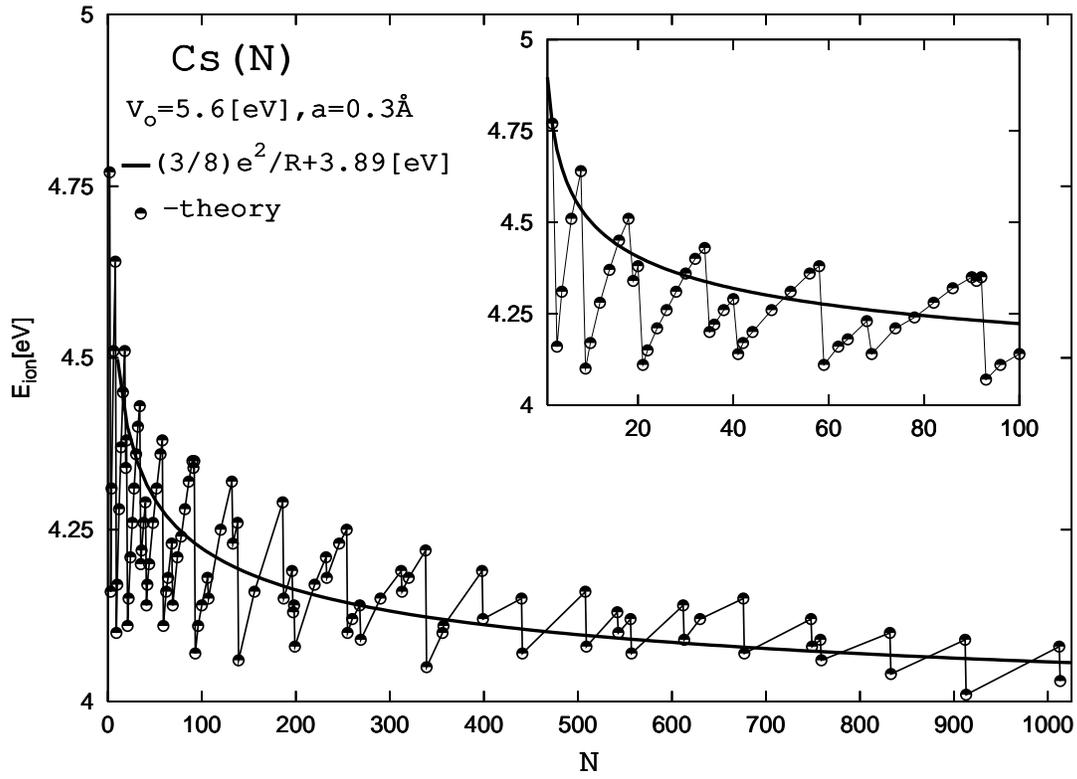}
\caption{The ionization energy of Cesium clusters Cs($N$).}
\end{figure}

\pagebreak[5]
\begin{figure}
\epsfxsize=160mm \epsfbox{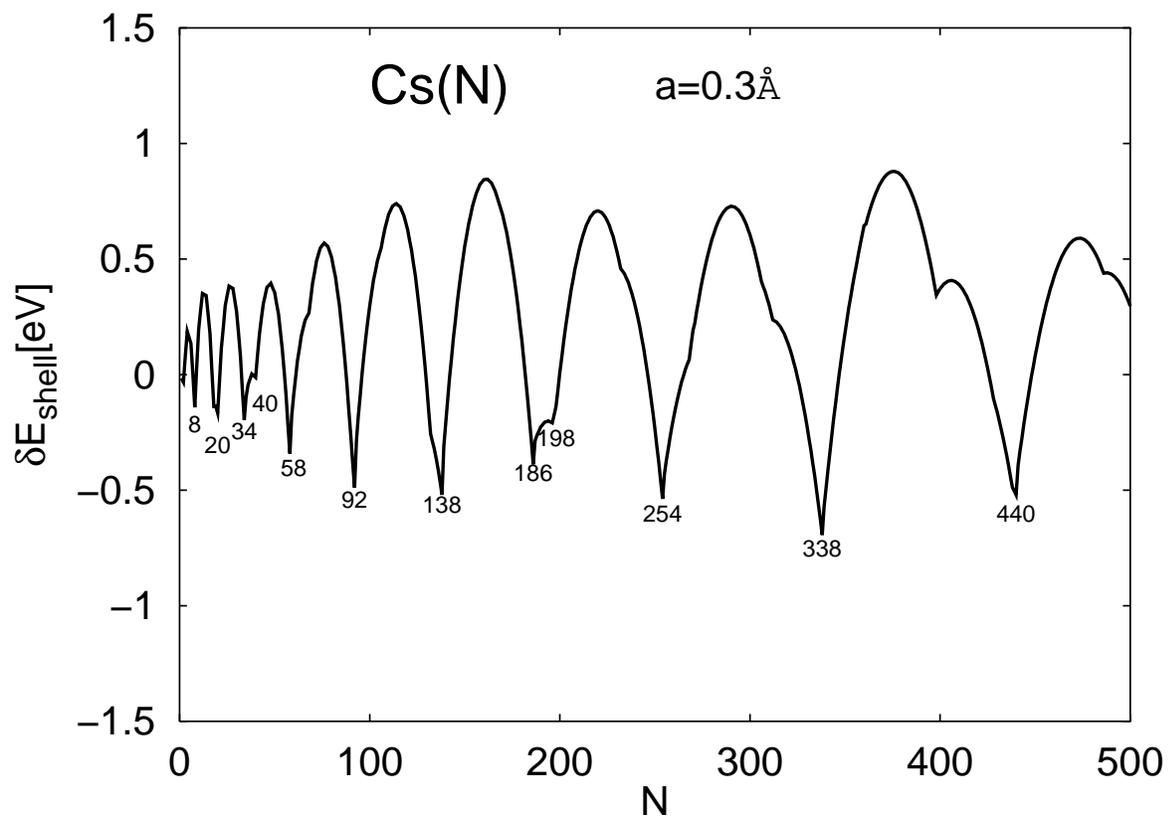}

\caption{The shell-correction energy and the magic numbers for the 
       Cs($N$) cluster evaluated with the surface thickness 
       $a$=0.3 {\AA}.} 
\end{figure}

\begin{figure}
\epsfxsize=160mm \epsfbox{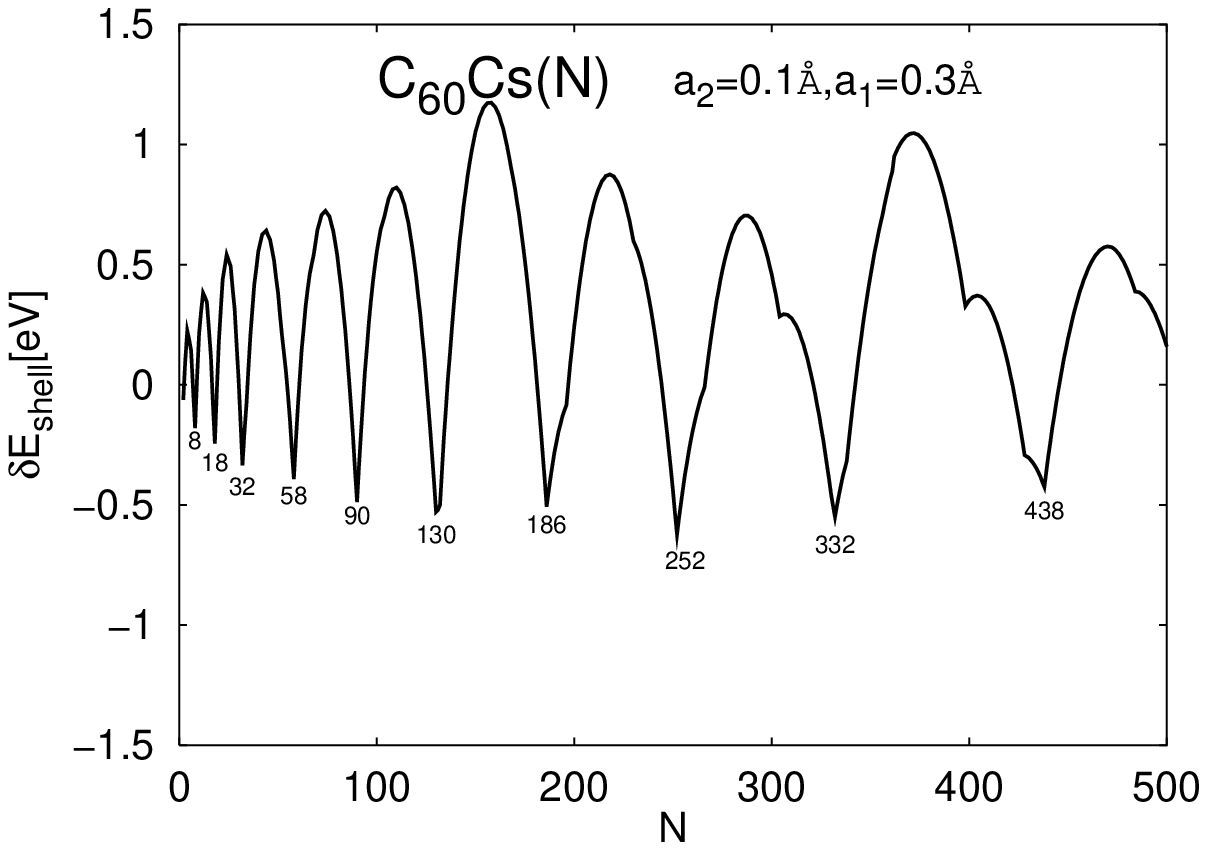}

\caption {Shell-correction energy and the magic numbers for the 
        C$_{60}$Cs($N$) cluster evaluated with the surface thickness 
        $a_1$=0.3 {\AA} and $a_2$=0.1 {\AA}.}

\end{figure}

\pagebreak[5]
\begin{figure}
\epsfxsize=150mm \epsfbox{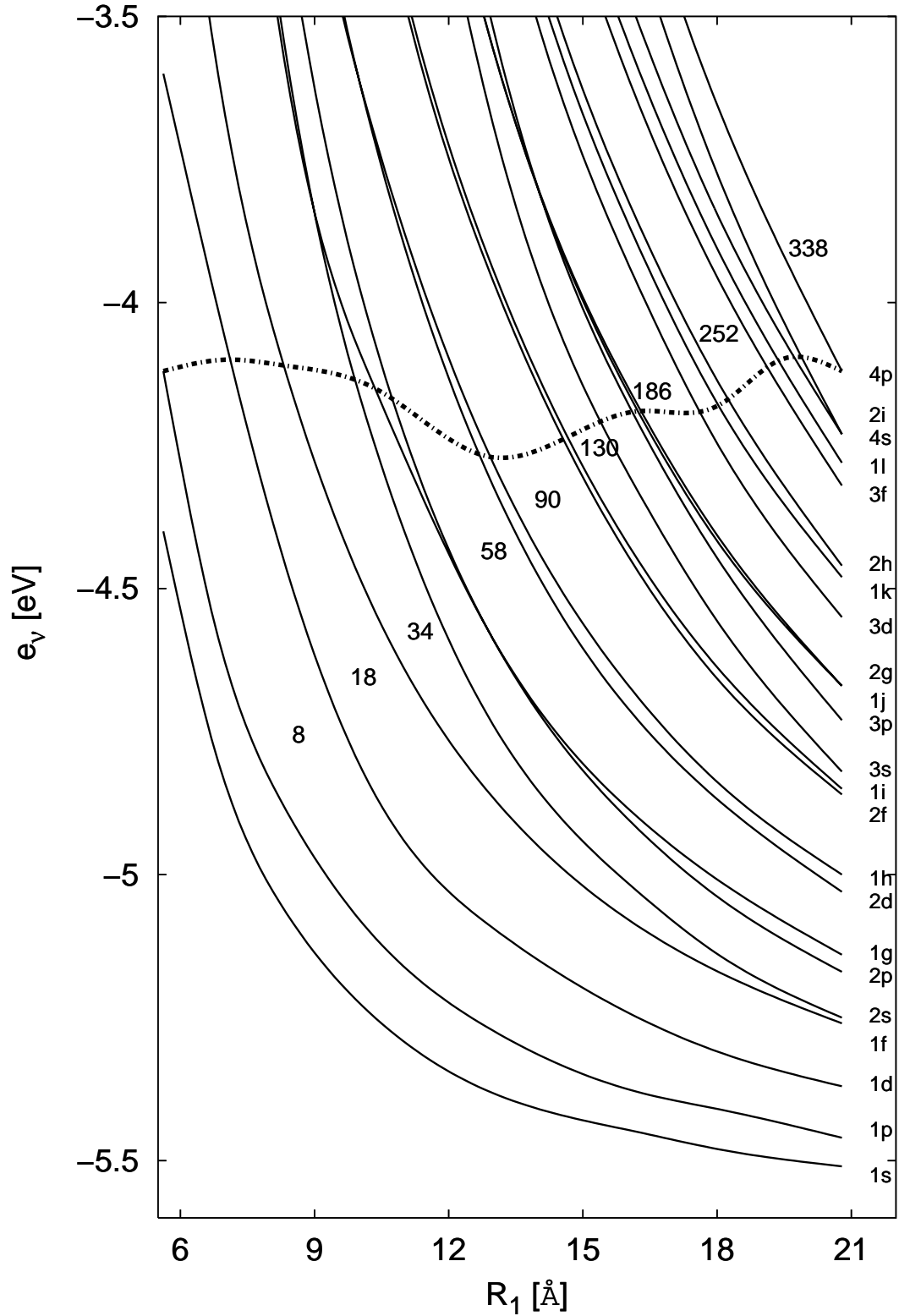}
\caption{The single-particle levels in the Cesium layer covering the spherical
      core of radius $R_2$. The depth of the single particle potential
      is equal to 5.6 eV, and the surface thicknesses $a_1$=0.3 {\AA} and 
      $a_2$=0.1 {\AA}. The black points show the position of the Fermi level.}

\end{figure}

\end{document}